\documentstyle[pre,aps,epsf]{revtex}
\input epsf
\newcommand{\be}{\begin{equation}}
\newcommand{\ee}{\end{equation}}
\newcommand{\bea}{\begin{eqnarray}}
\newcommand{\eea}{\end{eqnarray}}

\begin{document}
\draft
\title{Spin-spin interaction and magnetic state of 2-D Wigner crystal}

\author{V. V. Flambaum, I. V. Ponomarev$^{a}$, O.P.Sushkov}
\address{ School of Physics, The University of New South Wales,
Sydney 2052, Australia}

\maketitle

\begin{abstract}
It is demonstrated that there is anti-ferromagnetic spin-spin interaction
between nearest electrons in the 2D Wigner crystal. This is also valid for
the Wigner liquid - the state with destroyed long-range order but preserved
short-range one. We calculate the value of the anti-ferromagnetic interaction
(both analytically and numerically)
and discuss a possible magnetic state of the Wigner crystal. This state
can be strongly influenced by the spin-Peierls mechanism.
\end{abstract}


We consider a two dimensional electron gas ($1/r$ repulsion)
at zero temperature in the presence of a uniform neutralizing background.
It was shown a long  ago by Wigner \cite{Wig} that at sufficiently low
density the electron gas (or electron fluid) undergoes a transition into
a crystal state. This is because at low density the Coulomb interaction
dominates the kinetic energy and the correlated state becomes energetically
favorable. Analysis of the lattice dynamics shows that the stable
crystal structure in 2D is the triangular lattice \cite{lattice}.

The Wigner crystallization has been observed for electrons at the surface
of liquid helium \cite{helium}.
Another 2D system for which the electron density can be easily controlled
is an inversion layer at a semiconductor surface \cite{semic}.

Theoretically the Wigner crystallization has been studied using Monte Carlo
simulations, see e.g. Refs. \cite{Ceperley}. These calculations are pretty
reliable as far as the critical density is concerned. However there is some
controversy about possible Ferromagnetic Fermi liquid at a density
slightly higher than the crystallization density, see Refs.
\cite{Overhauser,Senatore}.

Interest in Wigner crystallization has been renewed recently after
observation of the insulator-conductor transition in dilute 2D
electron systems \cite{Krav1}.
Although this transition probably takes place in the liquid phase
it is pretty close to the point of crystallization.
A very interesting feature of the  transition is
suppression of the conducting phase by in-plane magnetic field \cite{Krav2}
which influences only spin degrees of freedom.

In the present work we calculate effective spin-spin interaction in 2D
Wigner crystal. This calculation is also valid for the Wigner liquid - 
the state with destroyed long-range order but preserved short-range order.

To avoid misunderstanding let us note that our calculation
does not show any magnetic phase transition in the liquid state
(i.e. there is no  ferromagnetic Fermi liquid between normal Fermi liquid and 
Wigner crystal). The state which we call Wigner liquid is just strongly 
renormalized  normal Fermi liquid.
Nevertheless magnetic properties of the Wigner liquid (and Wigner crystal)
 are quite unusual and 
somewhat similar to that of cuprate superconductors. There is competition
between superexchange (electron correlation) which gives antiferromagnetic
interaction between electron spins and the usual exchange Coulomb interaction
which gives ferromagnetic contribution. The superexchange is proportional to
$t^2/U$ where $t$ is the parameter which describes hopping of an electron to a
nearby site and $U$ is the Coulomb repulsion for two electrons sitting on the
same site. As a result, both the superexchange and exchange  are proportional
to the squared overlap between electron wave functions centered
on the different sites of the Wigner crystal . 
 Therefore,  simple estimates can not answer the question about the sign
of spin-spin interaction and we need more accurate calculations.
To provide better understanding and reliability of the results
we have performed these calculations twice: analytically and numerically.

Hamiltonian of the system under consideration is
\be\label{totH}
H =\sum_{i} {{{\bf p}_i^2}\over{2}} +
 \sum_{i<j}\frac{1}{|{\bf r_i}-{\bf r_j}|} + const,
\ee
where ${\bf p}_i$ and ${\bf r}_i$ are 2D momentum and coordinate respectively.
We use effective atomic units which means that all distances
are measured in units of the effective Bohr radius
$a_B^*=\hbar^2\epsilon/m^*e^2$, and energies in units of
${{m^*e^4}\over{\hbar^2\epsilon^2}}$. Here $m^*$ is the  
effective electron mass, and $\epsilon$ is the dielectric
constant which we assume to be independent of frequency.
Number density of the electrons $n$ is fixed by condition of
electroneutrality. An average distance $r_s$ between the electrons is
defined by $\pi r_s^2 =1/n$. It is well established \cite{Ceperley}
that the crystallization to the triangular lattice \cite{lattice}
occurs when $r_s \approx37$. In the presence of ``disorder'' further 
localization of the electrons stabilizes the Wigner crystal at higher
densities ($r_s\approx 10$, see \cite{Eguiluz}). The distance between the
nearest sites in the lattice is equal to $a=\sqrt{2\pi/\sqrt{3}}r_s 
\approx 1.90r_s$.

Electrostatic potential acting on the electron near equilibrium position
in the lattice is
\be\label{U}
U_1(r) \approx const+{{\gamma}\over{2}}{{r^2}\over{a^3}},
\ee
where $r\ll a$ is deviation from the equilibrium position.
To find $\gamma$ let us freeze all other electrons in their
equilibrium positions and calculate $U_1(r)$. Accounting for the six
nearest sites gives $\gamma=3$, and summation over entire lattice gives
\be\label{gam}
\gamma=3\sum_{n=1}^{\infty}\sum_{k=0}^{n-1}\frac{1}{(n^2+k^2-kn)^{3/2}}=5.5171.
\ee
Ground state electron wave function in the potential (\ref{U}) is
\be\label{psi}
\psi(r)={1\over{\sqrt{\pi}c}}e^{-r^2/2c^2}, \ \ \ c={{a^{3/4}}\over
{\gamma^{1/4}}}
\ee
In the above calculation we assume that size of the wave function is
much smaller than the lattice spacing, $c \ll a$, or $(a\gamma)^{1/4}\gg 1$.
For the crystallization point this parameter equals $(a\gamma)^{1/4}=4.4$,
and therefore the approximation is well justified in the crystal state.
We stress that the parameter appears in the exponent and therefore 4.4 is
a very large value.
Moreover, the approximation is justified in the liquid phase as soon as
$(a\gamma)^{1/4}\gg 1$. The matter is that the sum (\ref{gam}) is
saturated at 2-3 coordination circles and it is independent of the
presence or absence of the long-range order.
For conditions of the experiments \cite{Krav1,Krav2} this parameter
equals $(a\gamma)^{1/4}=3.1$.

 To find the magnitude of spin-spin interaction constant (the constant $J$,
which can be substituted to the Heisenberg Hamiltonian $J\sum_{\langle
i,j\rangle}\vec{S_i}\vec{S_j}$) we have to solve a two-particle problem,
freezing all the electrons except the nearest two ones which are shown by
crosses at Figure 1.
\begin{figure}\label{fig1}
\centerline{\epsffile{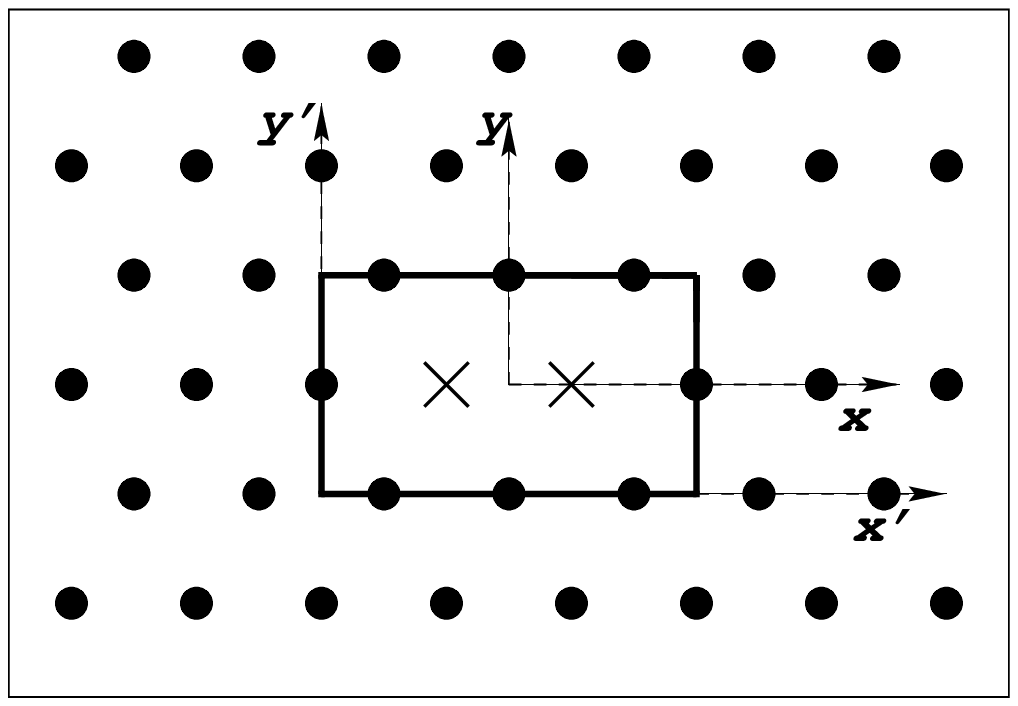}}
\caption{Boundary conditions for two-particles problem. }
\end{figure}

 Hamiltonian of the problem is
\be\label{2pham}
\hat{H}={{{\bf p}_1^2}\over{2}}+{{{\bf p}_2^2}\over{2}}
+U({\bf r_1}) +U({\bf r_2}) + \frac{1}{|{\bf r_1}-{\bf r_2}|},
\ee
where $U({\bf r})$ is potential of all frozen electrons
(dots at Fig. 1). The splitting between the  ground states for  total spin
$S=1$ and $S=0$ gives us the constant $J$:
\begin{eqnarray}
J &=&(E_g^{S=1}-E_g^{S=0})\  \equiv (E_A-E_S),\label{J}\\
\hat{H}\Psi_S &=& E_S\Psi_S,\label{eq1}\\
\hat{H}\Psi_A &=& E_A\Psi_A.\label{eq2}
\end{eqnarray}
Because of the Fermi statistics the two-electron wave function
 is antisymmetric with
respect to permutation. Therefore, the symmetric coordinate wave function
corresponds to spin $S=0$ and the antisymmetric one corresponds to $S=1$.

Of course the accurate solution of the problem can only be found 
(and have been found) numerically. However, to find the sign and
the basic dependence $J$ on a distance parameter $a$ we performed
an approximate analytical calculation. 
 To this end, we will follow the procedure, which was suggested
long time ago by Gorkov and Pitaevskii for the calculation of the term
splitting in hydrogen molecule\cite{Gorkov}.

We multiply  Eq.~(\ref{eq1}) by $\Psi_A$ and  Eq.~(\ref{eq2}) by $\Psi_S$,
take the difference between the results and calculate the integral over
some  region in configuration four dimensional space of the electrons.
We choose the integration volume in which $x_1\leq x_2$
(i.e. to the left of the hyperplane $\Sigma (x_1=x_2)$).
Using the Hamiltonian (\ref{2pham}) we obtain
\be\label{split1}
(E_S-E_A)\int\!\!\int_{\Omega}\Psi_A\Psi_S\, d{\bf r_1}\,
d{\bf r_2}=
\oint_{\Sigma}(\Psi_S\nabla \Psi_A -\Psi_A\nabla\Psi_S)\,{\bf d\Sigma}.
\ee
The kinetic energy term in the right-hand side is reduced to the surface
intergral using
$$\Psi_S\nabla^2 \Psi_A -\Psi_A\nabla^2\Psi_S=
\nabla (\Psi_S\nabla \Psi_A -\Psi_A\nabla\Psi_S)$$
and an integration by parts.

Now we introduce combinations of the functions $\Psi_{1,2}=1/\sqrt{2}
(\Psi_S\pm\Psi_A)$. They correspond to the states of "distinguishable"
particles, when, e.g. for $\Psi_1({\bf r_1, r_2})$, the first electron
is principally located near its equilibrium position $x=-a/2$
and the second electron near position $x=a/2$.
A simple calculation gives
$$\int\!\!\int_{\Omega}\Psi_S\Psi_A\, d{\bf r_1}\,
d{\bf r_2}=\frac{1}{2}\int\!\!\int_{\Omega}(\Psi_1^2-\Psi_2^2)
\, d{\bf r_1}\,d{\bf r_2}\approx 1/2.
$$
Substituting the wave functions $\Psi_{1,2}$ into Eq.~(\ref{split1})
and taking into account  that under ${\bf r_1} \leftrightarrow {\bf r_2}$
permutation  the wave functions  $\Psi_1 \leftrightarrow\Psi_{2}$, we obtain
\be\label{JvsInt}
J=-4\int\left [\Psi_{2}\frac{\partial\Psi_1}{\partial x_1}
\right]_{x_1=x_2}~dx_2~dy_1~dy_2.
\ee

 The formula (\ref{JvsInt}) shows that the main contribution to the
exchange constant is given by the region where the electrons are close to
each other. Indeed, the $x$ coordinates of both electrons coincide 
($x_1=x_2$), however, the $y$ coordinates may
be different. In this case there are strong correlations between the 
positions of the electrons due to Coulomb repulsion. This means that we
should go beyond the approximation where the two-particle wave function 
of the electrons is represented as a product of single-particles 
wave functions. 

It is easy to take into account the effect of the 
correlations in the  quadratic approximation.

 Assuming  that the particles are distinguishable and oscillate near their
equilibrium positions, we write the  Hamiltonian in the following form
\begin{eqnarray}\label{2phama1}
\hat{H}&=&-\frac{\Delta_1}{2}-\frac{\Delta_2}{2}
+ \frac{\omega^2}{2}\left ((x_1+a/2)^2+y_1^2
+ (x_2-a/2)^2+y_2^2\right)+\nonumber\\
& &+ \left\{ \frac{1}{|{\bf r_1}-{\bf r_2}|}
       -\frac{1}{|{\bf r_1}-{\bf a}/2|}
       -\frac{1}{|{\bf r_2}+{\bf a}/2|}\right\}.
\end{eqnarray}
Here the frequency $\omega=\sqrt{\gamma/a^3}$.

 The Hamiltonian (\ref{2phama1}) is valid at small displacements  $x_i$ and
$y_i$ from their equilibrium positions. Expanding the last term in the curly
brackets near ($\tilde x_{1,2}=x_{1,2}\mp a/2$) we finally get the following 
Hamiltonian in the quadratic approximation:
\be\label{2phama2}
\hat{H}=-\frac{\Delta_1}{2}-\frac{\Delta_2}{2}
+ \frac{\omega^2}{2}\left (\tilde x_1^2+y_1^2+\tilde x_2^2+y_2^2
-\frac{4}{\gamma} \tilde x_1\tilde x_2+
 \frac{2}{\gamma} y_1 y_2\right)+O(\tilde x^3/a^4).
\ee
Using an obvious change of  variables
\begin{eqnarray}\label{chvar}
u,v&=&\frac{y_1\pm y_2}{\sqrt{2}},\nonumber\\
\xi,\eta &=&\frac{ x_1 \pm  x_2}{\sqrt{2}},
\end{eqnarray}
we separate Hamiltonian (\ref{2phama2}) into four independent
oscillators with frequencies $\omega_{u,v}=\sqrt{(\gamma\pm 1)/a^3}$
and $\omega_{\xi,\eta}=\sqrt{(\gamma\mp 2)/a^3}$. Thus, the ground
state wave functions are
\begin{eqnarray}\label{grst}
\Psi_1(u,v,\xi,\eta) &= &
\frac{(\omega_u\omega_v\omega_{\xi}\omega_{\eta})^{1/4}}{\pi}\exp\left(-1/2
[
\omega_u u^2+\omega_v v^2+\omega_{\xi} \xi^2+\omega_{\eta}(\eta+a/\sqrt{2})^2
 ]\right),\nonumber\\
\Psi_2(u,v,\xi,\eta) &=&\Psi_1(u,v,\xi,-\eta).
\end{eqnarray} 
Substituting (\ref{grst}) in Eq.~(\ref{JvsInt}) we obtain
\be\label{Josin}
J=+2\omega_{\eta}a\int\left [
\Psi_2\Psi_1\right ]_{x_1=x_2}\, dx_2\, dy_1\, dy_2=
(\gamma+2)^{3/4}\sqrt{\frac{2}{\pi}}a^{-5/4}e^{-\sqrt{(\gamma+2)a}/2}.
\ee
This formula is presented in the atomic units.  In regular units it 
can be written as 
\be\label{Josreg}
J= +\frac{e^2}{\epsilon a}\left [\frac{a_B^*}{a}
\frac{4(\gamma+2)^3}{\pi^2}\right ]^{1/4}
\exp\left(-\sqrt{\frac{\gamma+2}{4}\frac{a}{a_B^*}}\right)
=3.62\frac{e^2}{\epsilon a}\left [\frac{a_B^*}{a}\right ]^{1/4}
\exp\left(-1.37\sqrt{\frac{a}{a_B^*}}\right)
\ee

The  plus sign in the exchange constant shows that the system is 
anti-ferromagnetic. It is worthwhile to note that the exponent 
in (\ref{Josin}) is different from $e^{-\sqrt{\gamma a}/2}$, which  
appears, if states $\Psi_{1,2}$ are represented by a product of independent
single-particle wave functions (\ref{psi}).

 In order to check the importance of  correlations and find the
correct exponent for $J$ we have performed numerical calculations of the
problem over a  rectangle area (see Fig. 1). To be absolutely correct we 
have to impose periodic boundary conditions in the rectangle.
However, the wave function is very small at the boundary and so
the results are not sensitive to the boundary condition.
It is much more convenient to make the wave function vanish at the boundary 
and this is the condition which we use in the present work.

The single particle basis set is given by (see Fig. 1)
\bea\label{bsf1e}
\phi_{\bf i}({\bf r}) \equiv \phi_{nm}(x',y')&=&
\frac{2}{\sqrt{AB}}\sin(\frac{\pi}{A}n x')\sin(\frac{\pi}{B}m y')\nonumber\\
\varepsilon_{\bf i} \equiv \varepsilon_{nm} &=&
{{\pi^2}\over{2}}\left[{{n^2}\over{A^2}}+{{m^2}\over{B^2}}\right],
\eea
where $A=3a$ and $B=\sqrt{3}a$.
Hence, for the two-electron problem the set is
\bea
&&|i\rangle\equiv |{\bf i_1}{\bf i_2}\rangle =
C_{{\bf i_1 i_2}}\left [\phi_{\bf i_1}({\bf r_1})\phi_{\bf i_2}({\bf r_2})
\pm \phi_{\bf i_1}({\bf r_2})\phi_{\bf i_2}({\bf r_1})\right ].\nonumber\\
&&E_i = \varepsilon_{\bf i_1}+\varepsilon_{\bf i_2}
\eea
The sign ``$+$'' corresponds to $S=0$ (anti-ferromagnetic), and the sign 
``$-$'' corresponds to $S=1$ (ferromagnetic). The normalization coefficient 
$C_{{\bf i_1 i_2}}=1/2$ if $\bf i_1 = \bf i_2$ otherwise it equals
$\frac{1}{\sqrt2}$. The matrix element of the Hamiltonian (\ref{2pham}) is 
of the form
\be\label{Hij}
\langle i|\hat{H}|j\rangle = E_i\delta_{ij}+
\langle i|\hat{V^{(1)}}|j\rangle +
\langle i|\hat{V^{(2)}}|j\rangle ,
\ee
where $\langle i|\hat{V^{(1,2)}}|j\rangle$ are matrix elements of
the single particle potential and the two-particle interaction correspondingly.
\bea
\langle i|\hat{V^{(1)}}|j\rangle &=&
2~C_{{\bf i_1 i_2}}C_{{\bf i_3 i_4}}\left [
V^{(1)}_{{\bf i_1 i_3}}\delta_{{\bf i_2 i_4}} \pm
V^{(1)}_{{\bf i_1 i_4}}\delta_{{\bf i_2 i_3}} \pm
V^{(1)}_{{\bf i_2 i_3}}\delta_{{\bf i_1 i_4}} +
V^{(1)}_{{\bf i_2 i_4}}\delta_{{\bf i_1 i_3}}
\right ]\nonumber\\
\langle i|\hat{V^{(2)}}|j\rangle &=&
2~C_{{\bf i_1 i_2}}C_{{\bf i_3 i_4}}\left [
V_{i_1 i_2 i_3 i_4}\pm V_{i_1 i_2 i_4 i_3} \right ].
\eea
To find $J$, which is exponentially small, we need a very large basis set.
The most time consuming part is the computation of the two-particle matrix
element $\langle i|\hat{V^{(2)}}|j\rangle$, which formally is a 4-dimensional
integral. Fortunately this integral can be reduced to an integral which
is effectively one dimensional. This reduction, which is demonstrated in the
Appendix, allowed us to perform computations with the size of Hilbert
space up to $N=1380$.

 Numerical solution of the problem was performed for two different cases.
Firstly, we considered  all the frozen electrons (points at Fig. 1) as 
point-like charges, which means that the mean-field potential in this case 
is just the sum of the Coulomb potentials:
\begin{eqnarray}\label{u_xy}
U(x,y)&=&\sum_{kl}u_0(|{\bf r_{kl}}-{\bf r}|),\nonumber\\
u_0(r)  &=&1/r.
\end{eqnarray}
The sum runs over sites of the triangular lattice.

Secondly, we considered  the density of the
frozen electrons to be distributed according to (\ref{psi}) and hence
\begin{eqnarray}\label{u_xy1}
U(x,y)&=&\sum_{kl}u_1(|{\bf r_{kl}}-{\bf r}|),\nonumber\\
u_1(r)=\frac{\sqrt{\pi}}{c}e^{-r^2/2c^2}I_0(r^2/2c^2),
\end{eqnarray}
where $I_0(x)$ is the modified Bessel function.
In both cases all the results  are very close and therefore we present
plots only for the second case.

The matrices (\ref{Hij}) for ferromagnetic and anti-ferromagnetic cases
were calculated and diagonalized. The Hilbert space was truncated at some
high energy state.
To be confident that the ground state was found with reasonable accuracy we
used two basis sets with dimensions $N=975$ and $N=1380$, where  $N$ is
the total number of two-particle states.

The plots of ground state electron density
\be
\rho({\bf r})=<0|\delta({\bf r}-{\bf r_1})+\delta({\bf r}-{\bf r_2})|0>
\ee
for  $a=45$, which corresponds to the limit of our calculations, 
are given in Figure  2. Similar plots for for $a=15$,
which corresponds to the conditions of the experiments \cite{Krav1,Krav2}, are 
given in Figure  3. The fact that the maximums coincide with the lattice sites 
tells us about the self-consistency of
the method. The shape of the density operator near the equilibrium positions
also corresponds to the expected Gaussian electron density, obtained from 
the combinations of the wave functions in (\ref{grst}): 
\begin{eqnarray}\label{rhoan}
\rho(x,y)_{S,A}&=&N_{S,A}\frac{2}{\pi}\sqrt{
\frac{\omega_u\omega_v\omega_{\xi}\omega_{\eta}}
{(\omega_u+\omega_v)(\omega_{\xi}+\omega_{\eta})}}e^{-\tilde\omega_y y^2}
 \left [
e^{-\tilde\omega_x (x+a/2)^2}+e^{-\tilde\omega_x (x-a/2)^2}\pm
2e^{-\omega_{\eta}a^2/2}e^{-\tilde\omega_x x^2}\right ].
\end{eqnarray}
Here $N_{S,A}=[1\pm e^{-\omega_{\eta}a^2/2}]^{-1}$ is the normalization 
coefficient due to the nonorthogonality of the functions $\Psi_1$ and 
$\Psi_2$, and 
$\tilde\omega_y=2\omega_u\omega_v/(\omega_u+\omega_v)$, 
$\tilde\omega_x=2\omega_{\xi}\omega_{\eta}/(\omega_{\xi}+\omega_{\eta})$.
\begin{figure}\label{fig2}
\epsfxsize=15 truecm
\centerline{\epsffile{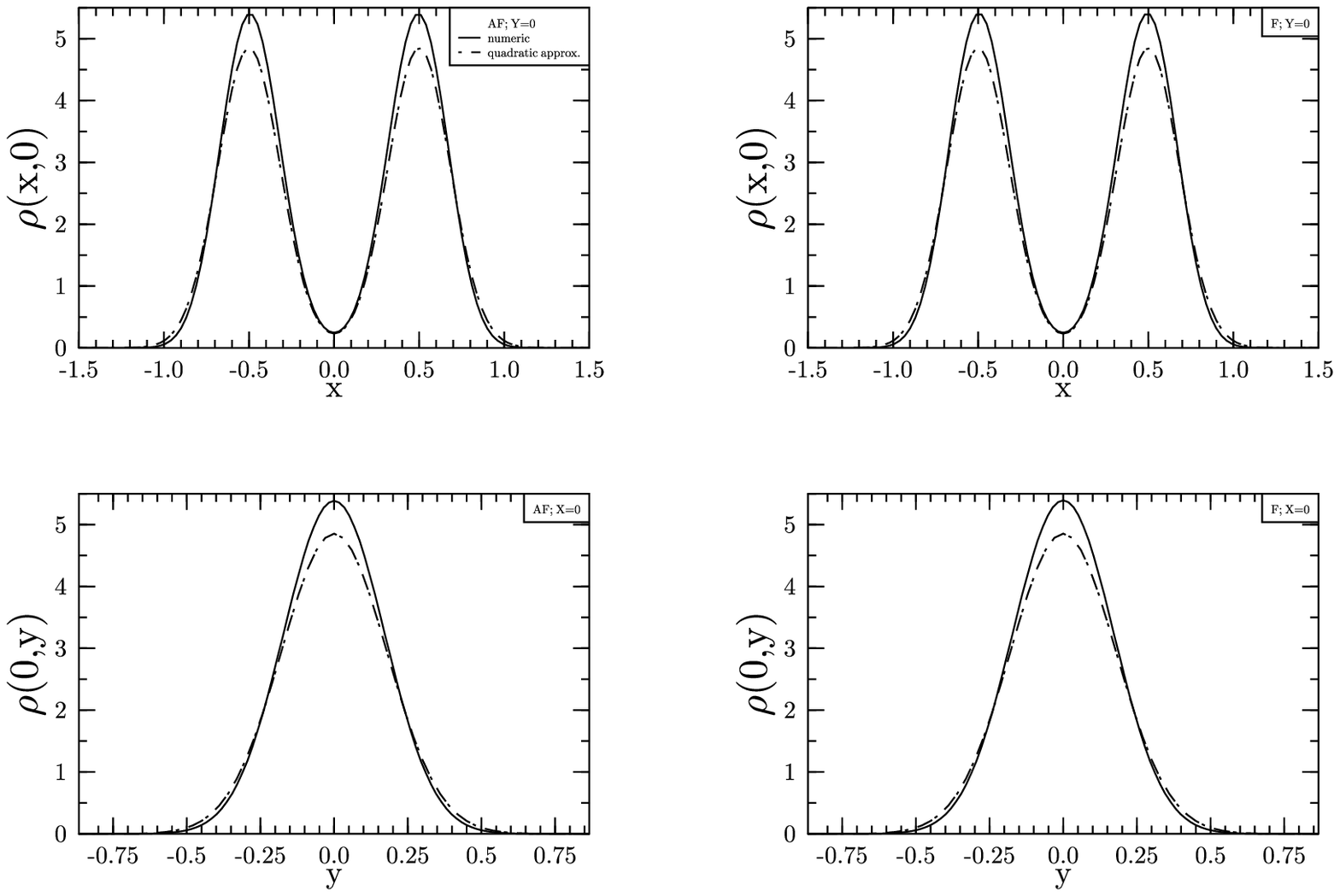}}
\caption{Profiles of the density operator $\rho(\vec{r})$ for the 
antiferromagnetic (two left plots) and ferromagnetic (right plots)
ground states for $a=45$ ($r_s=24$). The dot-dashed  lines show the 
analitical results in quadratic approximations ( formula ~(\ref{rhoan})). 
}
\end{figure}
\begin{figure}\label{fig3}
\epsfxsize=15 truecm
\centerline{\epsffile{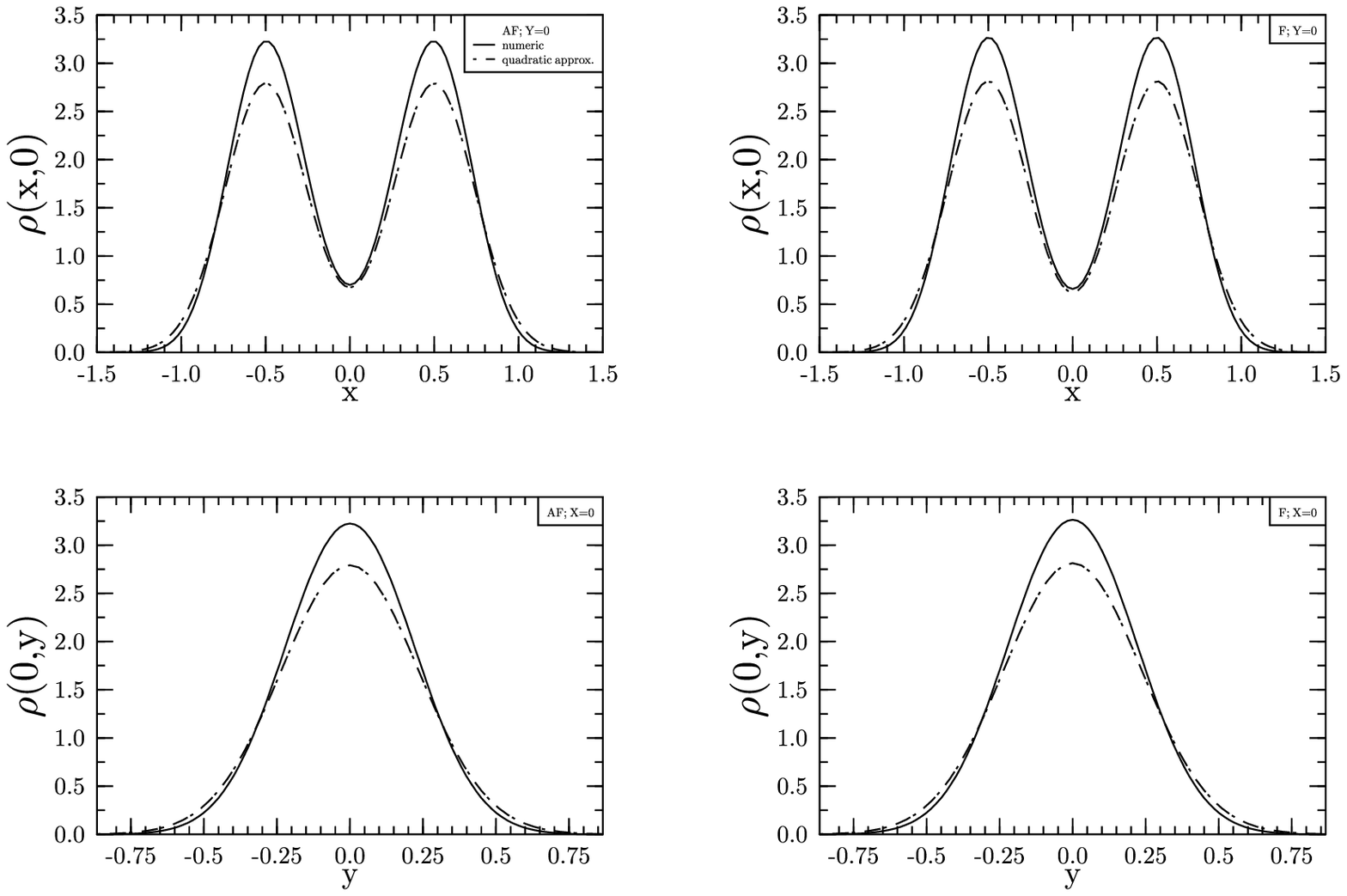}}
\caption{Profiles of the density operator $\rho(\vec{r})$ for 
the antiferromagnetic (two left plots) and ferromagnetic (right plots)
ground states for $a=15$ ($r_s=7.9$). The dot-dashed  lines show the 
analitical results in quadratic approximations ( formula ~(\ref{rhoan})).}
\end{figure}

We found that for the whole region of the strength parameter $a$ the 
anti-ferromagnetic ground state is always below the ferromagnetic one.
We obtained the following values of the constant $J(a)$ for the parameters 
of experiments \cite{Krav1,Krav2} ($\epsilon=8,\ m^*=0.19m_e$)
\begin{eqnarray}
J(15)&=&6.66\cdot 10^{-4} = 0.6 K,\nonumber\\
J(45)&=&1.48\cdot 10^{-6} = 1.4\cdot 10^{-3} K.\nonumber\\
\end{eqnarray}

Experiments \cite{Krav1,Krav2} correspond to $a \simeq 15$.
The behavior of $J$, as expected, has an exponential dependence 
$\sim e^{-\delta\sqrt{a}}.$
The plot of the dependence $\ln(J)$ vs. $\sqrt{a}$ summarizes our results 
in  Figure 4. Diamonds and crosses show the magnitude for different basis 
sets and nicely depict the truncation effects for large $a$ and 
for small number of basis states.
The dot-dashed line represents the theoretical curve (\ref{Josin}) and 
the solid line is the best fit.
\begin{figure}\label{fig4}
\epsfxsize=15 truecm
\epsfysize=6 truecm
\centerline{\epsffile{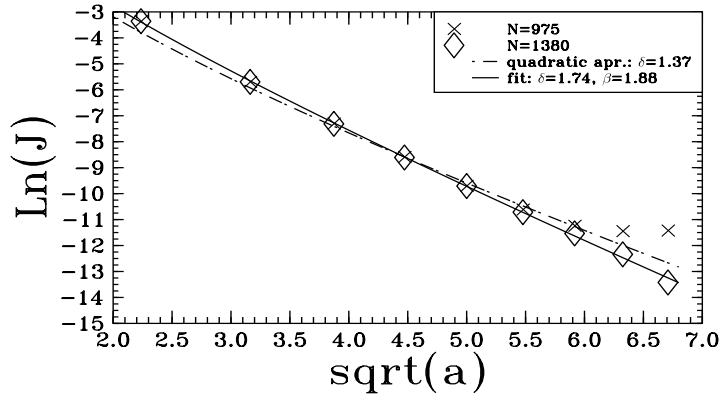}}
\caption{Dependence $\ln(J)$ on $\sqrt{a}$. Diamonds and crosses represent
data for large ($N=1380$) and shortened ($N=975$) bases correspondingly.
The dot-dashed  line shows the theoretical curve of Eq.~(\ref{Josin}).
The solid line is the best fit by curve $y=C-\delta\sqrt{a}-\beta\ln(a)$.
}
\end{figure}

 We fitted  our data by two different functions.
For the first one we fixed the power of $a$ in the preexponential
factor (the analytical formula (\ref{Josin}) gives $a^{-5/4}$):
$$\ln(J)=C-\delta\sqrt{a}-5/2\ln(\sqrt{a})$$
 and got $C=2.25, \delta=1.595$.
Let us  note that the constant $\delta$ is slightly larger than the
predicted analytical value $\delta_{an}=\sqrt{\gamma+2}/2=1.37$.
This fact shows that  the higher order terms in the expansion of the 
Hamiltonian (\ref{2phama1}) are important for getting the correct
magnitude of the exponent.

 In the second case we looked for the best parameters for
$$\ln(J)=C-\delta\sqrt{a}-\beta\ln(\sqrt{a}).$$
 We obtained $C=2.02, \delta=1.74, \beta=1.88$.

  The analytical formula for J in Eq.~(\ref{Josin}) and more accurate
fits of the numerical calculation data allows one to estimate the value of 
$J$ for the Wigner crystal and Wigner liquid states in the large
region of densities.

Due to geometric frustration  collinear long-range antiferromagnetic
order is, strictly speaking, not possible on a triangular lattice.
The possible solution in this case
is a system spin wave function which in the zero approximation consists of  
spin zero pairs. The antiferromagnetic interaction increases
when the distance between the electrons decreases. Therefore, there should be
a tendency for nearby electrons coupled to spin zero pairs to move slightly
toward each other. This phenomenon is usually called the ``Spin-Peierls''
mechanism.

  It is interesting that the value of the spin-spin interaction $J=J(15)$
is comparable with the energy $\mu H$ in the experiments  \cite{Krav1,Krav2},
where $H$ is the critical magnetic field destroying conductivity.
We can speculate that this field  effectively transforms the
system to a ferromagnetic state. In the ferromagnetic state the conductivity
should be smaller than in the antiferromagnetic state. Indeed, hopping of the
electrons with opposite spins from one site to another is allowed by the Pauli
principle. The magnetic field rearranges spins
in the same direction. In this case such hopping is suppressed
by Pauli blocking. This possibly  destroys the conductivity.


\acknowledgments
We are grateful to M. Kuchiev, L.~\'Swierkowski, D.~Neilson and
D. Shepelyansky for useful discussions.
V.V.F. is grateful to the Special Research Center For the Subatomic Structure
of Matter , University of Adelaide where part of this work has been done.
This work is supported by Australian Research Council.

\appendix

\section{Matrix elements of the interaction}
The eigenfunctions and eigenvalues for the single-electron problem in the
two-dimensional rectangle with sides $A=3a$ and $B=\sqrt{3}a$ are
given by Eq.~(\ref{bsf1e}).
The matrix element of the Coulomb interaction between two unfrozen electrons is
given by the integral over the  rectangle's area
\be\label{me2d1}
V_{i_1 i_2 i_3 i_4} = \int
\phi_{\bf i_3}({\bf r_1'})^*\phi_{\bf i_4}({\bf r_2'})^*\frac{1}
{|{\bf r_1'}-{\bf r_2'}|}\phi_{\bf i_1}({\bf r_1'})\phi_{\bf i_2}({\bf r_2'})
 d^2r_1' d^2r_2'
\ee
It is convenient to change variables (see Fig. 1):
\be
u=x_1'-x_2',\qquad v=y_1'-y_2',\qquad x_2''=x_2',\qquad  y_2''=y_2'
\ee
Introducing notations
$\tilde u ={{\pi u}\over{A}}$, $\tilde v ={{\pi v}\over{B}}$, and
\bea\label{ras1}
F_{n_1 n_2 n_3 n_4}(\tilde x_1,\tilde x_2)&=&4\sin(\tilde x_1 n_1)
\sin(\tilde x_1 n_3)\sin(\tilde x_2 n_2)\sin(\tilde x_2 n_4),
\nonumber\\
W_{n_1 n_2 n_3 n_4}(\tilde u)&=&
\int_0^{\pi-\tilde u}F_{n_1 n_2 n_3 n_4}
({\tilde x_2}+\tilde u,{\tilde x_2}) d {\tilde x_2}\nonumber\\
J_{n_1 n_2 n_3 n_4}&=& \frac{1}{2}\left (1+ (-1)^{n_1+n_2+n_3+n_4}\right ),
\eea
the matrix element (\ref{me2d1}) can be rewritten as:
\be\label{me2d2}
V_{i_1 i_2 i_3 i_4}=J_{n_1 n_2 n_3 n_4}J_{m_1 m_2 m_3 m_4}
\frac{4}{AB\pi^2}\int_0^A\!\int_0^B
W_{n_1 n_2 n_3 n_4}(\tilde u)W_{m_1 m_2 m_3 m_4}(\tilde v)
\frac{du\,dv}{\sqrt{u^2+v^2}}.
\ee
In this transformation we use the following relation
$$\int_0^{\pi-\tilde u}F_{n_1 n_2 n_3 n_4}(x, x+\tilde u)\, dx=
(-1)^{n_1+n_2+n_3+n_4}
\int_0^{\pi- \tilde u}F_{n_1 n_2 n_3 n_4}(x+\tilde u, x)\, dx$$.

 It is convenient to calculate the double integral (\ref{me2d2}) using
polar coordinates
$\tilde u ={{\pi u}\over{A}}={{r}\over{\sqrt{3}}}\cos t$,
$\tilde v ={{\pi v}\over{B}}=r \sin t$. Taking into account that
$tan^{-1}(B/A)=\pi/6$ we find that
\bea\label{ras}
V_{i_1 i_2 i_3 i_4}&=&J_{n_1 n_2 n_3 n_4}J_{m_1 m_2 m_3 m_4}
\frac{4}{3\pi^3}\left [V_{\bf 1} + V_{\bf 2}\right ]\nonumber\\
V_{\bf 1}&=& \int_0^{\pi/6}\,dt\!\int_0^{\pi\sqrt{3}/\cos(t)}
W_{n_1 n_2 n_3 n_4}(r\cos(t)/\sqrt{3})W_{m_1 m_2 m_3 m_4}(r\sin(t))\, dr,
\nonumber\\
V_{\bf 2}&=& \int_0^{\pi/3}\,dt\!\int_0^{\pi/\cos(t)}
W_{n_1 n_2 n_3 n_4}(r\sin(t)/\sqrt{3})W_{m_1 m_2 m_3 m_4}(r\cos(t))\, dr.
\eea
In order to write down an analytical expression for the function
$W_{n_1 n_2 n_3 n_4}(\tilde u)$ let us introduce the following
notations
\be
\left\{ \begin{array}{ll}
n=|n_3 - n_1|, & n=0,1,2,\ldots\\
m=|n_4 - n_2|, & m=0,1,2,\ldots\\
l= n_3 + n_1 , & l=2,3,\ldots\\
k= n_4 + n_2 , & k=2,3,\ldots\\
\end{array} \right.
\ee
Using (\ref{ras1}) one can find that
\be
W_{n_1 n_2 n_3 n_4}(\tilde u)= \left\{\begin{array}{ll}
f_1(n,m,l,k;\tilde u) &\\
f_2(n,l,k;\tilde u) & \mbox{if }\; n=m\neq 0\\
f_2(l,n,m;\tilde u) & \mbox{if }\; l=k\\
-f_2(n,m,l;\tilde u) & \mbox{if }\; n=k\\
-f_2(m,n,k;\tilde u) & \mbox{if }\; m=l\\
f_3(l,k;\tilde u) & \mbox{if }\; n=m=0\\
f_4(n,l;\tilde u) & \mbox{if }\; n=m\neq 0 \mbox{ and } l=k\\
f_5(l;\tilde u) & \mbox{if }\; n=m=0 \mbox{ and } l=k.\\
\end{array} \right.
\ee
where
\bea
f_1(n,m,l,k;u)&=&{\frac {n\sin(nu)\left ({k}^{2}-{m}^{2}\right )}
{\left ({n}^{2}-{k}^{2
}\right )\left ({n}^{2}-{m}^{2}\right )}}+{\frac {m\sin(um)\left ({l}^
{2}-{n}^{2}\right )}{\left ({m}^{2}-{l}^{2}\right )\left ({m}^{2}-{n}
^{2}\right )}}-{\frac {l\sin(lu)\left ({k}^{2}-{m}^{2}\right )}{\left
({l}^{2}-{m}^{2}\right )\left ({l}^{2}-{k}^{2}\right )}}-
{\frac {k\sin
(uk)\left ({l}^{2}-{n}^{2}\right )}{\left (-{n}^{2}+{k}^{2}\right )
\left ({k}^{2}-{l}^{2}\right )}}\nonumber\\
f_2(n,m,l;u)&=&
\frac{\pi -u}{2}\cos(nu)+\frac{\sin(nu)}{2n}
\left (1+2\,{\frac {{n}^{4}-{m}^{2}{l}^{2}}{\left ({n}^{2}-{m}^{2}\right )
\left ({n}^{2}-{l}^{2}\right )}}\right )
+{\frac {m\sin(um)\left ({n}^{2}-{l}^{2}\right )}
{\left ({m}^{2}-{l}^{2}\right )\left ({m}^{2}-{n}^{2}\right )}}
+{\frac {l\sin(lu)\left ({n}^{2}-{m}^{2}\right )}
{\left ({l}^{2}-{n}^{2}\right )\left ({l}^{2}-{m}^{2}\right )}}\nonumber\\
f_3(n,m;u)&=&
\pi -u-{\frac {\sin(um){n}^{2}}{\left ({m}^{2}-{n}^{2}\right )m}}-{
\frac {\sin(nu){m}^{2}}{\left ({n}^{2}-{m}^{2}\right )n}}\nonumber\\
f_4(n,m;u)&=&
\frac{\pi -u}{2}\left [\cos(nu)+\cos(um)\right ]+
\frac {\sin(nu)}{2n}\frac{3\,{n}^{2}+{m}^{2}}{{n}^{2}-{m}^{2}}+
\frac {\sin(um)}{2m}\frac{3\,{m}^{2}+{n}^{2}}{{m}^{2}-{n}^{2}}
\nonumber\\
f_5(n;u)&=&
\pi -u+\frac{\pi -u}{2}\,\cos(nu)+\, \frac{3}{2n}\sin(nu)
\eea
This completes the description of the calculation procedure for the
two-electron Coulomb matrix element. The advantage is that
each of the two integrals in (\ref{ras}) require numerical work
equivalent only to the computation of a $1D$ integral.

 Calculation of the the single-particle matrix element of the external
potential $U({\bf r})$ is much simpler. It is convenient to use
$x$ and $y$ instead of $x'$ and $y'$
(see Fig. 1). Then
\bea
V^{(1)}_{{\bf i_1 i_3}} &\equiv & V^{(1)}_{\{n_1 m_1,n_3 m_3\}} =
\frac{4}{AB}(-1)^{\frac{n+m}{2}}J_{n_1 n_3}J_{m_1 m_3} 
\int_0^{A/2}\int_0^{B/2} U(x,y)\times \nonumber\\
& &\times
\left\{\cos(\frac{\pi}{A}nx) - (-1)^{n_2}\cos(\frac{\pi}{A}lx)\right\}
\left\{\cos(\frac{\pi}{B}my) - (-1)^{m_2}\cos(\frac{\pi}{B}ky)\right\}
dx\, dy,
\eea
where
\be
\left\{ \begin{array}{lll}
n &=& |n_3 - n_1|\\
m &=& |m_3 - m_1|\\
l &=& n_3 + n_1 \\
k &=& m_3 + m_1
\end{array} \right.
\ee


\end{document}